\documentclass[finals,epj,nopacs,twocolumn]{svjour}

\usepackage{latexsym}
\def\makeheadbox{\relax}

  \newcommand{\Qa}{\mathcal{Q}}
\usepackage{graphics}
\usepackage{hyperref}
\usepackage{mathtools}
\usepackage{amsmath}
\usepackage{cuted}
 \usepackage[latin1]{inputenc}
\usepackage{color}
\usepackage{url}
\usepackage{graphicx}
\usepackage{float}
\usepackage{amsmath}
\usepackage{tikz}
\usepackage{graphicx}
\usepackage{latexsym,color}
\usepackage{amssymb}
\usepackage{xcolor,color}
\usepackage{amsfonts,dsfont}

\usepackage[T1]{fontenc}
\usepackage{ae,aecompl,latexsym}

\definecolor{blue}{rgb}{0.0,0.0,0.75}
\definecolor{purple}{rgb}{0.78,0.0,0.75}
\hypersetup{linkcolor=purple,citecolor=blue,filecolor=cyan,urlcolor=magenta}

\def\be{\begin{equation}}
\def\ee{\end{equation}}
\def\bea{\begin{eqnarray}}

\def\eea{\end{eqnarray}}

\newcommand{\la}{\mathcal{A}}
\newcommand{\Pa}{\mathcal{P}}
\newcommand{\laa}{\mathcal{L}}
\usepackage{physics}

\newcommand{\il}{~}

\usepackage[varg]{txfonts}

\begin{document}
\title{Extracting information on  black hole  horizons}

\author{Daniela  Pugliese\inst{1} \and Hernando Quevedo\inst{2}
}                     
\institute{ $^1$ Research Centre for Theoretical Physics and Astrophysics,
	Institute of Physics,
	Silesian University in Opava,
	Bezru\v{c}ovo n\'{a}m\v{e}st\'{i} 13, CZ-74601 Opava, Czech Republic
	\\     $^{2}$
	Instituto de Ciencias Nucleares, Universidad Nacional Aut\'onoma de M\'exico,  AP 70543, M\'exico, DF 04510, Mexico \\
		Dipartimento di Fisica and ICRA, Universit\`a di Roma ``La Sapienza", I-00185 Roma, Italy \\
	Department of Theoretical and Nuclear Physics,
	Kazakh National University,
	Almaty 050040, Kazakhstan
}
\abstract{
We present some  features of Kerr black hole  horizons that are replicated on  orbits accessible to outside observers.
We use the concepts of horizon confinement and replicas to show that outside the outer horizon there exist  photon orbits whose frequencies contain information about the inner horizon and that can, in principle, be detected through the emission spectra of black holes.
It is shown that such photon orbits exist close to the rotation axis of the Kerr geometry.
We argue that these results could be used to recognize and further investigate black holes and their horizons.
}
\date{Received: date / Revised version: date}
\maketitle

\section{Introduction}\label{Sec:impe-null-i}

The horizon is a paramount feature of black holes (BHs) entering in   numerous  astrophysical processes and in the understanding of the  physics and geometry  bordering  quantum gravity.
 The {Event Horizon Telescope}\footnote{\href{}{https://eventhorizontelescope.org/}.} (EHT)   has produced  in 2019 the first  BH image, the  black hole's shadow, as a bright ring--like structure with a  central dark disk\cite{ETH1}.
 More recently, the {EHT} collaboration has shown the presence of a   polarised  fraction of light in  the M87 galaxy,  close to the BH boundary,  measuring for the first time  light polarisation which is interpreted as a sign of magnetic fields presence \cite{ETH2,ETH3}. This detection opens a very rich perspective for the analysis of matter and fields subject to  BH gravitational stresses, particularly in the jets emission.

In this letter, we study  the properties of BH horizons in  the Kerr geometry by  investigating photon orbits with orbital frequency (relativistic velocity) equal in magnitude  to the horizons frequencies.
The possible detection of these photons would provide information on the properties of the horizons. Such photon orbits are shown to exist especially in regions  close to the BH rotational axis.
We  interpret the results in the  more general framework of  metric bundles (MBs), introduced in \cite{observers,remnants,Pugliese:2019efv,Pugliese:2019rfz,LQG-bundles,PS,bundle-EPJC-complete} and connected to the structures considered in
 \cite{de-Felice-first-Kerr,de-Felice3,de-Felice-mass,de-Felice4-overspinning}, and \cite{Chakraborty:2016mhx,Tanatarov:2016mcs} and
\cite{Mukherjee:2018cbu,Zaslavskii:2018kix,Zaslavskii:2019pdc}.

In \cite{bundle-EPJC-complete}  in the wider  context of the  BHs and naked singularities correspondence, the  metric bundles structure  was  studied in detail, considering metric bundles as curves in a given plane called the extended plane. In that context it was highlighted the possibility that a part of these structures evidences  the presence of replicas  purposed for  the   analysis of photon circular motion  and  the determination of some characteristics of the inner and outer BH   horizons particularly close to the rotation axis.
Here in Sec.\il(\ref{Sec:pri-photon-fre})  we provide the analysis and complete classification of the  co-rotating and counter-rotating replicas of the horizons of the Kerr metric in the  black holes spacetimes, then focusing on  the spacial cases of the equatorial plane and the case of extreme  Kerr BH.
In Sec.\il(\ref{Sec:bundle-description}) we discuss the results in the bundles frame, interpreting our results  through these structures, commenting  on  the relevance of the results and their phenomenological consequences.

\textbf{The Kerr geometry}
The Kerr geometry is an  exact,  asymptotically flat, vacuum solution of the Einstein equations,    for an axisymmetric,   stationary spacetime,
with
(ADM) mass parameter $M\geq0$, rotational parameter (spin---specific angular momentum) $a\equiv J/M\in [0,M]$, where   $J$ is the
total angular momentum. For  $a=0$,  the spacetime is    the limiting static and spherically symmetric  Schwarzschild  geometry.
For $a=M$ the solution is known as Kerr extreme BH, for $a>M$ there are naked singularity solutions.
For the purposes of this work, it is convenient to represent the Kerr geometry in Boyer-Lindquist coordinates as
%
\bea\label{Eq:metric-1covector}&&
\dd s^2=-\alpha^2 \dd t^2+\frac{\Pa \sigma}{\Sigma} (\dd \phi- \omega_{zamos} \dd t)^2+\frac{\Sigma}{\Delta}\dd r^2+\Sigma \dd\theta^2,
\eea
where
\bea
&&\nonumber
\Delta\equiv r^2-2Mr+a^2,\quad\mbox{and}\quad\Sigma\equiv r^2+a^2(1-\sigma) ,
\\&&\nonumber  \Pa \equiv (r^2+a^2)^2-a^2 \Delta \sigma,\quad \mbox{with}\quad\sigma\equiv\sin^2\theta,
\eea
where
$\alpha=\sqrt{\Delta \Sigma/\Pa }$  is  the lapse function  and $\omega_{zamos}=2 a M r/\Pa $ is the  frequency of the zero angular momentum observer (ZAMOS). 
(In the following analysis, to simplify the discussion, when not otherwise specified we  use geometrical units with $r\rightarrow r/M$ and $a\rightarrow a/M$.).

The inner and outer  BH horizons   are located at
\bea
r_{\mp}=M\mp\sqrt{M^2-a^2},
\eea
and the  inner and outer ergosurfaces  in  the {Kerr} geometry are  the radii
\bea
r_{\epsilon}^{\mp}=M\mp\sqrt{M^2-a^2 (1-\sigma)}\ ,
\eea
respectively.
We consider  null-like
circular orbits with  rotational frequencies     $\omega_{\pm}:{g}(\mathcal{L},{\mathcal{L}})=0$, defined from  the  Killing vector $\mathcal{L}=\xi^t +\omega \xi^{\phi}$,   in terms of the
stationary $\xi^{t}=\partial_{t} $  and
axisymmetric $\xi^{\phi}=\partial_{\phi} $
     Killing vector fields.
 The  limiting   frequencies (or relativistic velocities) $ \omega_\pm$:
\bea
\omega_\pm\equiv\frac{2 a r  \pm\sqrt{\Delta  \Sigma^2/ \sigma }}{\Pa }\ ,
\eea
 are also the limiting frequencies bounding the
 (time-like) stationary observers   four-velocity.
 In this letter we use the relation  $\omega_\pm(r_{\pm})=\omega_H^{\pm}$, where  $\omega_H^{\pm}$ are the frequencies of the outer and inner Kerr horizons respectively. The null vector fields  $\mathcal{L}_H^{\pm}\equiv \mathcal{L}(r_{\pm})=\xi^t +\omega_H^{\pm} \xi^{\phi}$ in fact define the
horizons of the Kerr BH  as   Killing horizons (being generators of Killing event  horizons). (For $a=0$, the horizon  of the  Schwarzschild BH  is a Killing horizon with respect to the
Killing field
$\xi^t$, consequently   the
event, apparent, and Killing horizons   coincide. ).

Quantity $\omega_H^{\pm}$, expressing the BH rigid rotation, regulates  (with the BH surface gravity)  the BH thermodynamic laws and
The variation of the   BH irreducible  mass
$
\delta M_{irr}\geq 0$  constrained  by $ (\delta M-\delta J \omega^+_H)\geq 0 $ for a variation of mass and momentum.


\section{Horizon replicas: photon  frequencies  as horizon frequencies}\label{Sec:pri-photon-fre}
We study   the  solutions of  $g(\mathcal{L},\mathcal{L})=0$ associated to orbits with radius $r$, which are  different from the horizon radii,  but corresponding  to a photon orbit with  orbital frequency     equal  in magnitude to the frequency $\omega_H^{\pm}$ of the BH horizons, we define these orbits as horizons replicas.
 We  consider the corotating,  $a\omega>0$ and  the counter-rotating  $a\omega<0$  replicas with   frequencies that
are equal in magnitude  to the horizon frequencies. Notice that
$g(\mathcal{L},\mathcal{L})(a,-\omega)=g(\mathcal{L},\mathcal{L})(-a,\omega)$ and thus
$g(\mathcal{L},\mathcal{L})(-a,-\omega)=g(\mathcal{L},\mathcal{L})(a,\omega)$.

Replicas  therefore connect  two null vectors, $ \mathcal{L}(r_q,a,\sigma_q)$ and
$ \mathcal{L}(r_p,a,\sigma_p)$, where $r_q\neq r_p$ is  an  outer   or inner Killing horizon and in general $\sigma_p\neq\sigma_q$.
 We analyze the two regions $r<r_-$ (inner region) and $r>r_+$ (outer region). It turns out that solutions  exists   on planes  close to the BH rotation axis.
To represent the solutions, it is convenient to introduce the planes $(\sigma_v,\sigma_z)$ and spins $(a_{l},a_s)$ that are defined  Eqs.\il(\ref{Eq:sigmav},\ref{Eq:sigmaz},\ref{Eq:al},\ref{Eq:as}) of the Appendix
and represented in Fig.\il(\ref{FIG:ARESE1Pola1}). We introduce also the replicas $(r_n,r_z)$ that are the solutions described in  Eqs.\il(\ref{Eq:rn}) and (\ref{Eq:rz}). Moreover, we use the notation  $r_{\Qa}^{\mathbf{I}}$ and $r_{\Qa}^{\mathbf{II}}$ to designate one and two replicas, respectively, which are solutions of the multi-parametric equation for the  radius $r_{\Qa}\in \{r_n,r_z\}$.

The inequality $\omega_{+}>\omega_-$ is valid in general, except on the horizons.
In the
case $\omega_+=-\omega_H^\pm$,  there are no solutions.
More generally we classify the solutions as follows.

\textbf{Corotating inner horizon replicas  with $\omega_+=\omega_H^-$: }
\bea
&&r<r_-: \checkmark  \\\nonumber
&&r>r_+:
\sigma\in [0,\sigma_{crit}],\left(
 a=\{a_s,1\},  r^{\mathbf{I}}_z\right);
\left(a\in ]a_s,1[,r^{\mathbf{II}}_z\right).
\eea
 with $\sigma_{crit}=2 \left(2-\sqrt{3}\right)\approx 0.536$. Alternatively,
\bea&&
r>r_+: (a\in ]0,1[,  \sigma\in ]0,\sigma_{crit}[, r^{\mathbf{II}}_z),\\\nonumber
&&\qquad \qquad(a\in ]0,1[, \sigma=\sigma_{crit}); (a=1, \sigma\in ]0,\sigma_{crit}[), r^{\mathbf{I}}_z.
\eea
\textbf{Corotating  inner horizons replicas with
$\omega_-=\omega_H^-$:}
\bea&& r>r_+: \checkmark.\\
&&\nonumber r<r_-:
\sigma\in ]\sigma_{crit},1[:
(a=a_s, r=r^{\mathbf{I}}_z);\left(
a\in ]a_s,1[ , r=r^{\mathbf{II}}_z\right)
\\
&&\nonumber
\sigma\in ]\sigma_{crit},1[: \left(a=1,
r^{\mathbf{I}}_z\right),\sigma=1: \left(a\in ]0,1[,
r^{\mathbf{I}}_z\right).
\eea
Alternatively,
\bea&&
a\in ]0,1[: (\sigma_z,r^{\mathbf{I}}_z); \left(
\sigma\in ]\sigma_z,1[, r=r^{\mathbf{II}}_z\right),
\\&&\nonumber
 \left(a\in ]0,1[,\sigma=1\right);
\left(a=1,
\sigma\in [\sigma_{crit},1]\right), \quad r=r^{\mathbf{I}}_z.
\eea
\textbf{Corotating outer  horizons replicas with $\omega_+=\omega_H^+$: }
\bea
&&r\in [0,r_-[: \checkmark
\\\nonumber
&&r>r_+:
 \left(a\in ]0,1[, \sigma\in ]0,1]\right); 
\left(a=1, \sigma\in ]0,\sigma_{crit}[\right) r=r_z^{\mathbf{I}}.
\eea
\textbf{Corotating outer horizon replicas with $\omega_-=\omega_H^+$:}
\bea&&
r>r_+: \checkmark
\\
&&\nonumber r\in [0,r_-[:
(a\in ]0,1[, \sigma \in ]0,1[),
(a=1, \sigma\in ]\sigma_{crit},1[), r=r^{\mathbf{I}}_z
\eea
\textbf{Counter-rotating  inner horizons replicas with  $
\omega_-=-\omega_H^-$:}

Solutions cannot be in the ergoreigon,
  $r\in  [r_{\epsilon}^-,r_-[\cup]r_+,r_{\epsilon}^+]$.
Then:
\bea&&\label{Eq:counter-mmm}
r\in ]0,r_{\epsilon}^-[:(a\in ]0,1], \sigma\in ]0,1[, r^{\mathbf{I}}_n)
\\\nonumber
&&
r>r_{\epsilon}^+:
\sigma]0,\sigma_k[,  (a_l, r^{\mathbf{I}}_n),\left(
a\in ]a_l,1] , r^{\mathbf{II}}_n\right),
\\
&&\nonumber\qquad \qquad
\sigma=\sigma_k,  \left(a=1, r^{\mathbf{I}}_n\right),
\eea
in this case $a\in ]0,1]$ and {$\sigma_k=2 \left(8-3 \sqrt{7}\right)$}. Alternatively,
\bea&&\label{Eq:counter-mmm1}
r>r_{\epsilon}^+:
a\in ]0,1[, (\sigma\in ]0,\sigma_v[, r_n^{\mathbf{II}});
(\sigma_v, r_n^{\mathbf{I}})
\\&&\nonumber
a=1: (\sigma\in ]0,\sigma_k[, r_n^{\textbf{II}}); (\sigma_k,r_n^{\mathbf{I}}).
\eea
\textbf{Counter-rotating  outer horizons replicas with $\omega_-=-\omega_H^+$}
\bea&&\label{Eq:counter-mmm2}
r\in ]0,r_{\epsilon}^-[: (a\in ]0,1],\sigma\in ]0,1[, r=r^{\mathbf{I}}_n)
\\\nonumber
&&r>r_{\epsilon}^+: \sigma\in ]0,\sigma_k[: \left(a\in ]0,1],r^{\mathbf{II}}_n\right),
\\\nonumber
&&
 \sigma=\sigma_k:  \left( a\in ]0,1[,r^{\mathbf{II}}_n\right);
  (a=1,r^{\mathbf{I}}_n).
\\&&\nonumber
 \sigma\in ]\sigma_k,1]:
 \left(a\in ]0,a_l[, r^{\mathbf{II}}_n\right);
 \left(a=a_l,  r^{\mathbf{I}}_n\right).
\eea
 Alternatively,
\bea&&\label{Eq:counter-mmm3}
r>r_{\epsilon}^+:
 a\in ]0,a_{crit}[:  \left(\sigma\in ]0,1],
r^{\mathbf{II}}_n\right),
\\\nonumber
&&
 a=a_{crit}:
 \left(\sigma\in ]0,1[,r^{\mathbf{II}}_n\right);
\left(\sigma=1,r^{\mathbf{I}}_n\right)
 \\\nonumber
 &&
 a\in ]a_{crit},1[:\left(
 \sigma\in ]0,\sigma_v], r^{\mathbf{II}}_n\right);
 \left(\sigma=\sigma_v,
r^{\mathbf{I}}_n\right)
\\&&\nonumber
 a=1: \left(\sigma \in ]0,\sigma_k[,r^{\mathbf{II}}_n\right); \left(\sigma=\sigma_k, r^{\mathbf{I}}_n\right),
 \eea
 with   $a_{crit}\equiv 0.5784M$.

We consider now the particular case of an  extreme Kerr BH, $a=M$, and the situation on the equatorial plane  $\sigma=1$. Then,

\textbf{Extreme Kerr BH:}
%
%
 \bea&&
\omega_-=\omega_H^\pm:
(r>r_+,\checkmark);
(r<r_-,\sigma\in ]\sigma_{crit},1[,r_z^{\mathbf{I}}).
\\\label{Eq:counter-mmm4}
&&
\omega_-=-\omega_H^-:
 r>r_{\epsilon}^+,(\sigma\in ]0,\sigma_k[, r_{n}^{\mathbf{II}});  (\sigma_k,r_{n}^{\mathbf{I}}).
\\
&&\nonumber
 \quad\quad\quad\quad (r\in ]0,r_{\epsilon}^-[,\sigma\in ]0,1[,r_z^{\mathbf{I}}).
\\\label{Eq:counter-mmm5}
 &&
\omega_-=-\omega_H^+: (r>r_{\epsilon}^+,\sigma\in ]0,\sigma_k[,r_z^{\mathbf{II}});
 (\sigma_k,r_z^{\mathbf{I}}).
 \\
 &&\nonumber\quad\quad\quad\quad
\left(r\in]0,r_{\epsilon}^-[, \sigma\in ]0,1[,  r_z^{\mathbf{I}},\right)
\\
&&\omega_+=\omega_H^\pm:(r<r_-,\checkmark); \left(r>r_+, \sigma\in ]0,\sigma_{crit}[,r_z^{\mathbf{I}}\right)
\eea
(Note the cases $\omega_{(\pm)}=\omega_H^{\pm}$ and $\omega_-=-\omega_H^{\pm}$).

\textbf{The equatorial plane $\sigma=1$:}

In \cite{remnants} it was discussed the existence of  two corotating  orbits
$r_\pm^\pm$ such that $\omega_*(r_\pm^\pm)=\omega_H^\pm$, respectively, and
$r_-^-<r_-<r_+<r_+^+$ (almost everywhere but at  $a_g=0$ for the static spacetime and  $a_g=M$ for the extreme Kerr BH).
More in general here we find,
\bea\label{Eq:invt}&& \omega_-=\omega_H^-: (r>r_+,\checkmark); (r\in [0,r_-[,a\in ]0,1[, r_z^\mathbf{I}).
\\
&&\label{Eq:counter-mmm6}
\omega_-=\omega_H^+: \checkmark,\quad
\omega_-=-\omega_H^-: \checkmark.
\\\nonumber
&&\omega_-=-\omega_H^+: (r<r_{\epsilon}^-,\checkmark);(r>r_{\epsilon}^+,a\in ]0,a_{crit}[,r_n^{\mathbf{II}});
(a_{crit},r_n^\mathbf{I})
\\\label{Eq:counter-mmm7}
&&\omega_+=\pm\omega_H^-: \checkmark;\quad
\omega_+=-\omega_H^+:  \checkmark
\\\label{Eq:counter-mmm7a}
&&\omega_+=\omega_H^+:  (r>r_+,\checkmark);(r\in ]0,r_-[,
a\in ]0,1[, r_z^{\mathbf{I}})
\eea
(note the cases $\omega_\pm=\omega_H^\pm$).
We summarize the situation as follows.

\textbf{Inner horizon counter-rotating replicas with $\omega= -\omega_H^-$:}
\bea\label{Eq:counter-mmm8}&&\omega= -\omega_H^-:
(r\in [0,r_{\epsilon}^-[,a\in ]0,1], \sigma\in ]0,1[,r^{\mathbf{I}}_n).
\\&&\nonumber\qquad
r>r_{\epsilon}^+: (\sigma\in ]0,\sigma_k[,a_l,r_n^\mathbf{I}),
\\\nonumber
&&\qquad \qquad
(a\in ]a_l,1],r_n^{\mathbf{II}}),
(\sigma=\sigma_k,a=1,r_n^\mathbf{I}).
\eea
\textbf{Inner horizon corotating replicas with $\omega=\omega_H^-$:}
\bea&&\label{Eq:counter-mmm9}\omega=\omega_H^-: r>r_+,\sigma\in ]0,\sigma_{crit}[:\quad (a=a_s,r_z^\mathbf{I}),
\\\nonumber
&&\qquad \qquad(a\in ]a_s,1[,r_z^{\mathbf{II}});\quad(a=1,r_z^\mathbf{I}).
\\\nonumber
&&
r\in [0,r_-[, \sigma\in ]\sigma_{crit},1[; (a_s,r_z^\mathbf{I});(a\in ]a_s,1[,
r_z^{\mathbf{II}});
\\
&&\nonumber\qquad \qquad
(a=1,\sigma=1,r_z^\mathbf{I}); (\sigma=1,a\in]0,1[,
r_z^\mathbf{I}).
\eea
\textbf{Outer horizon counter-rotating replicas with $\omega=-\omega_H^+$:}
\bea&&\label{Eq:counter-mmm10}\omega=-\omega_H^+:(r\in [0,r_{\epsilon}^-[
,a\in ]0,1],\sigma\in ]0,1[,r_{n}^\mathbf{I})
\\
&&\nonumber
r>r_{\epsilon}^+:
(\sigma\in ]0,\sigma_k[,a\in ]0,1],r_n^{\mathbf{II}});
\\\nonumber
&&\sigma=
\sigma_k: (a\in ]0,1[, r_n^{\mathbf{II}});
(a=1, r_n^{\mathbf{I}});
\\\nonumber
&&
\sigma\in ]\sigma_k,1]:
(a\in ]0,a_l[,r_n^{\mathbf{II}}); (a=a_l; r_n^{\mathbf{I}})
\eea
\textbf{Outer horizon  corotating replicas with $\omega=+\omega_H^+$:}
\bea&&
\omega=\omega_H^+:  \quad r=r^{\textbf{I}}_{z}
\\\nonumber
&&r\in [0,r_-[:(\sigma\in ]0,1[,a\in ]0,1[); (a=1,\sigma\in ]\sigma_{crit},1[),
\\&& \nonumber r>r_+:(a\in ]0,1],\sigma\in ]0,\sigma_{crit}[); (a\in ]0,1[,\sigma\in [\sigma_{crit},1]).
\eea
Examples of replicas are shown in Fig. \il(\ref{FIG:ARESE1Pola1}).
To clarify this concept   in Fig.\il(\ref{FIG:plot4sit7}),  we show a set of replicas of the inner and outer horizons.  A study of the asymptotic region is illustrated also in   Fig.\il(\ref{FIG:ARESE1Pola1}). As clear from Figs\il(\ref{FIG:ARESE1Pola1}), there are two  extreme values of
  $\theta$ for the existence of the inner horizon replicas  in the outer region.
 A further interesting aspect is the asymptotic behavior (for large $r/M$),  where the outer horizon  and the curves representing the  inner horizon replicas close and approach the BH poles (i.e., $\theta=\{0,\pi\}$, that is, for larger $r$ and small $\sigma$).
For $r \rightarrow + \infty$ there is $\omega_ {\pm} =
  0 $.  
 This means that at a fixed
   angle $\sigma$,   the inner  horizon  replica approaches  the outer horizon  replica. Similarly, at a fixed radius  $r/M$ and for values of  $\theta$ approaching the BH poles, there are two replicas for two  horizon frequencies, respectively.    We note that for
   small  $\sigma$  and large  $r$ the two curves, inner and outer horizons replicas, get closer.
%
%
\begin{figure*}
    \includegraphics[ width=4cm]{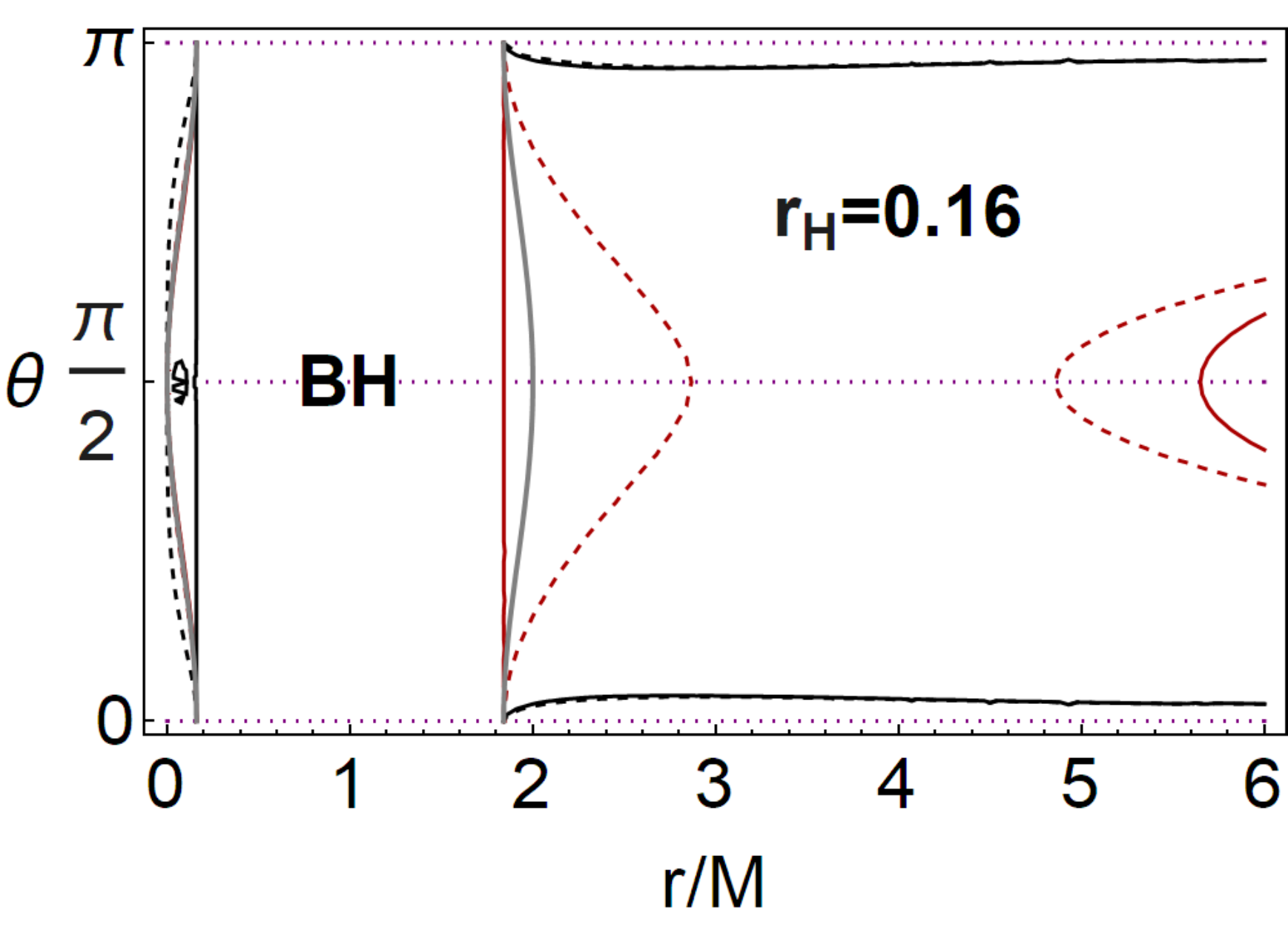}
    \includegraphics[width=4cm]{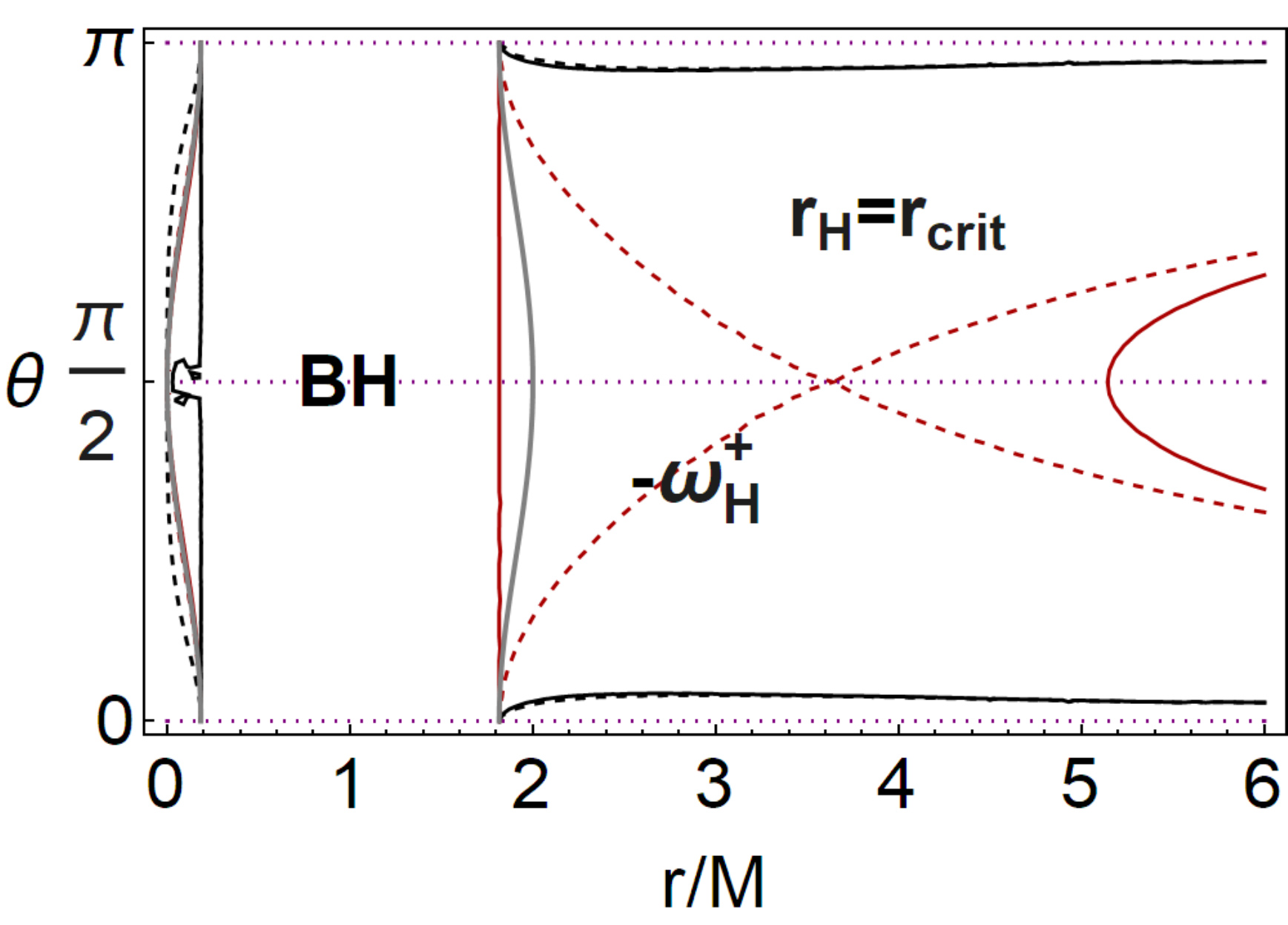}  \includegraphics[width=4cm]{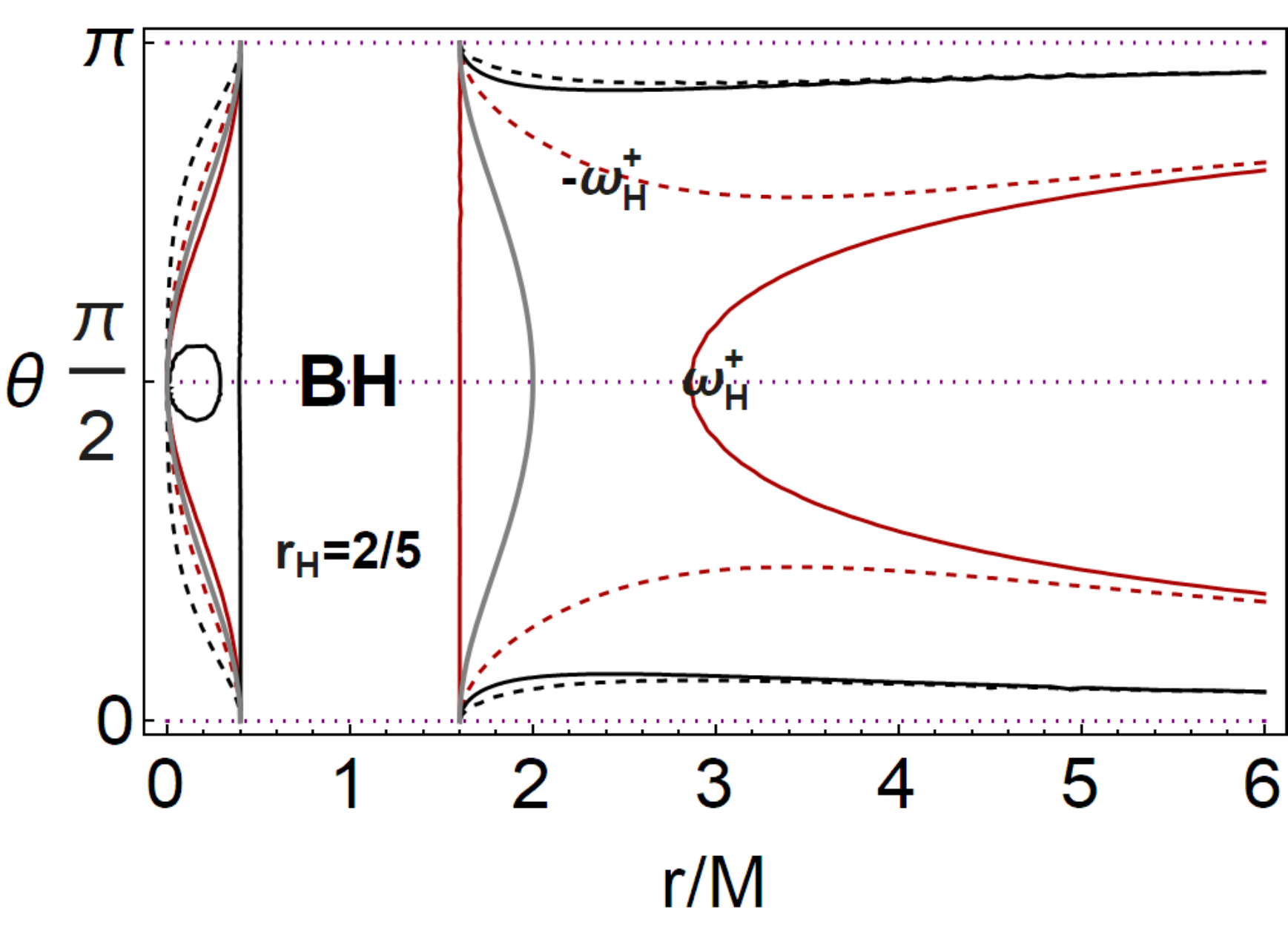}
    \includegraphics[width=4cm]{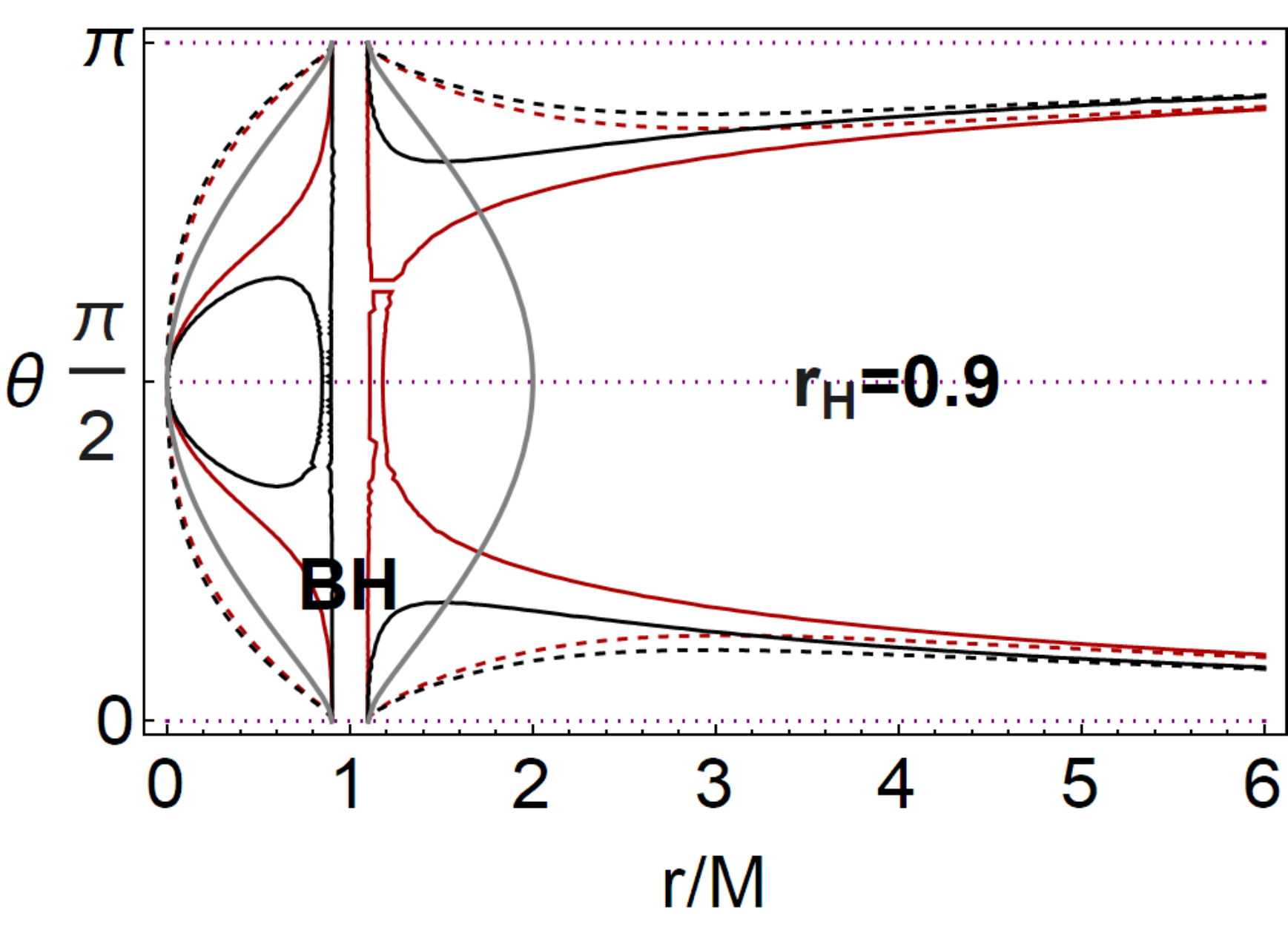}
\includegraphics[width=4cm]{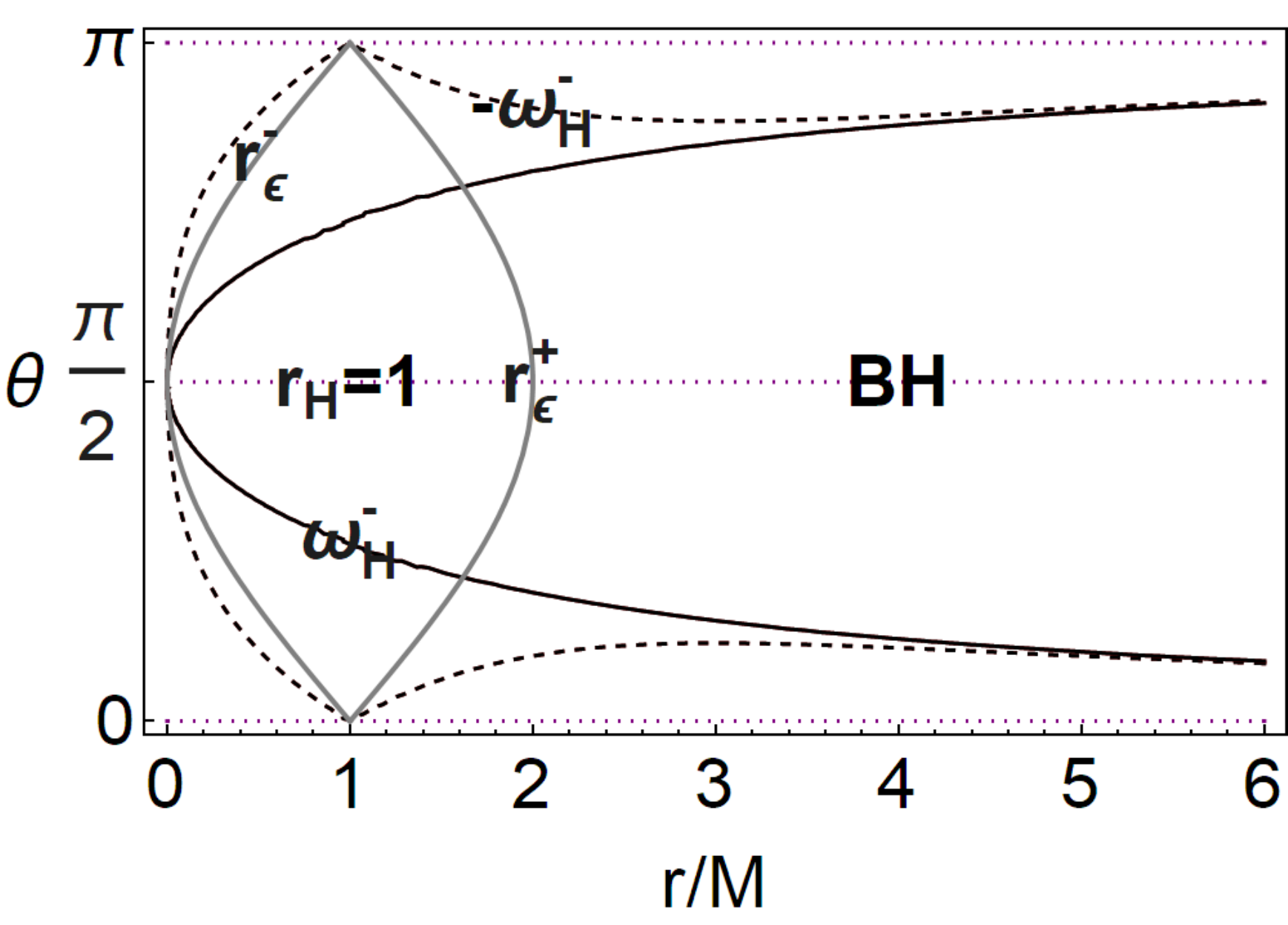}
 \includegraphics[width=4.cm]{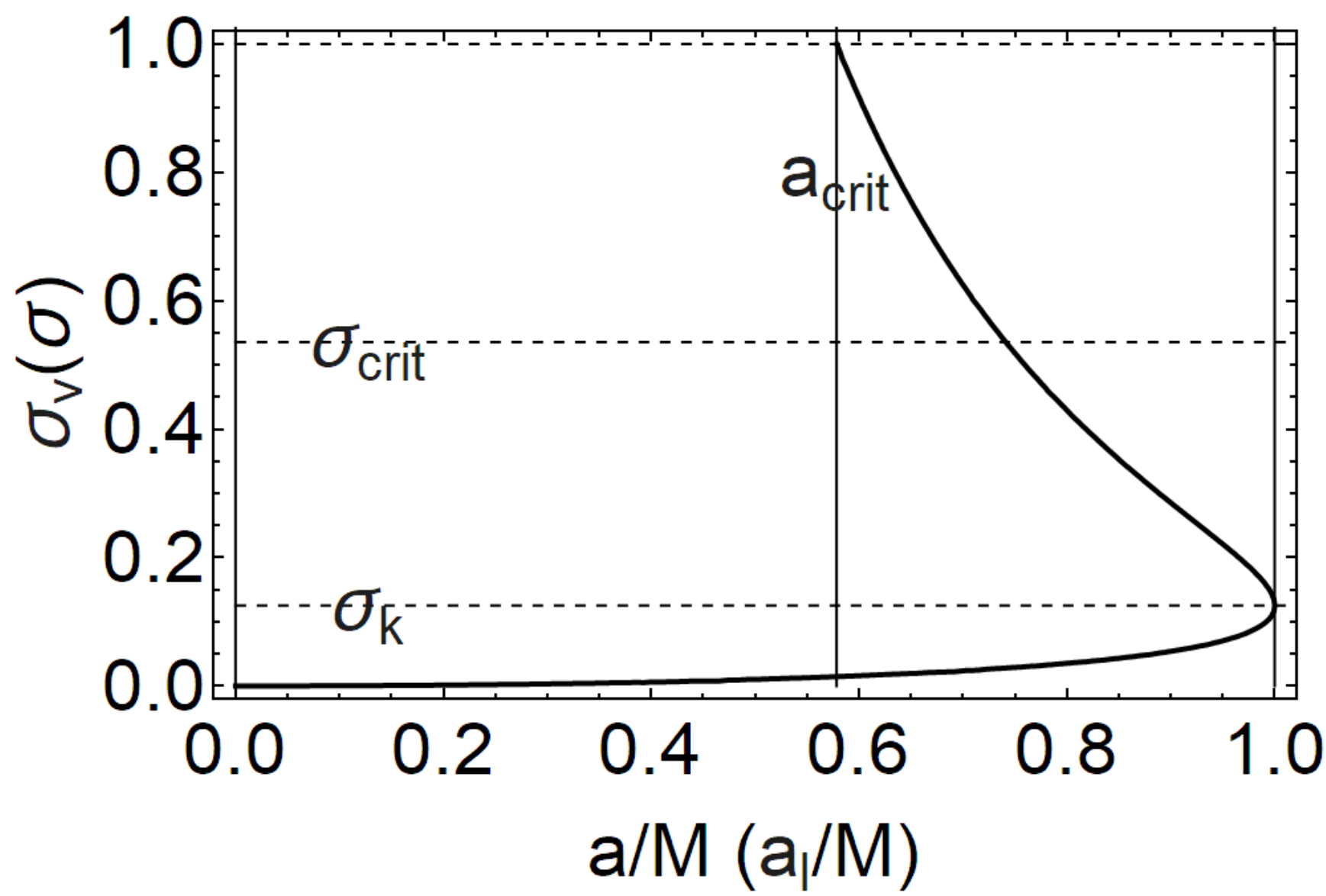}
          \includegraphics[width=4.cm]{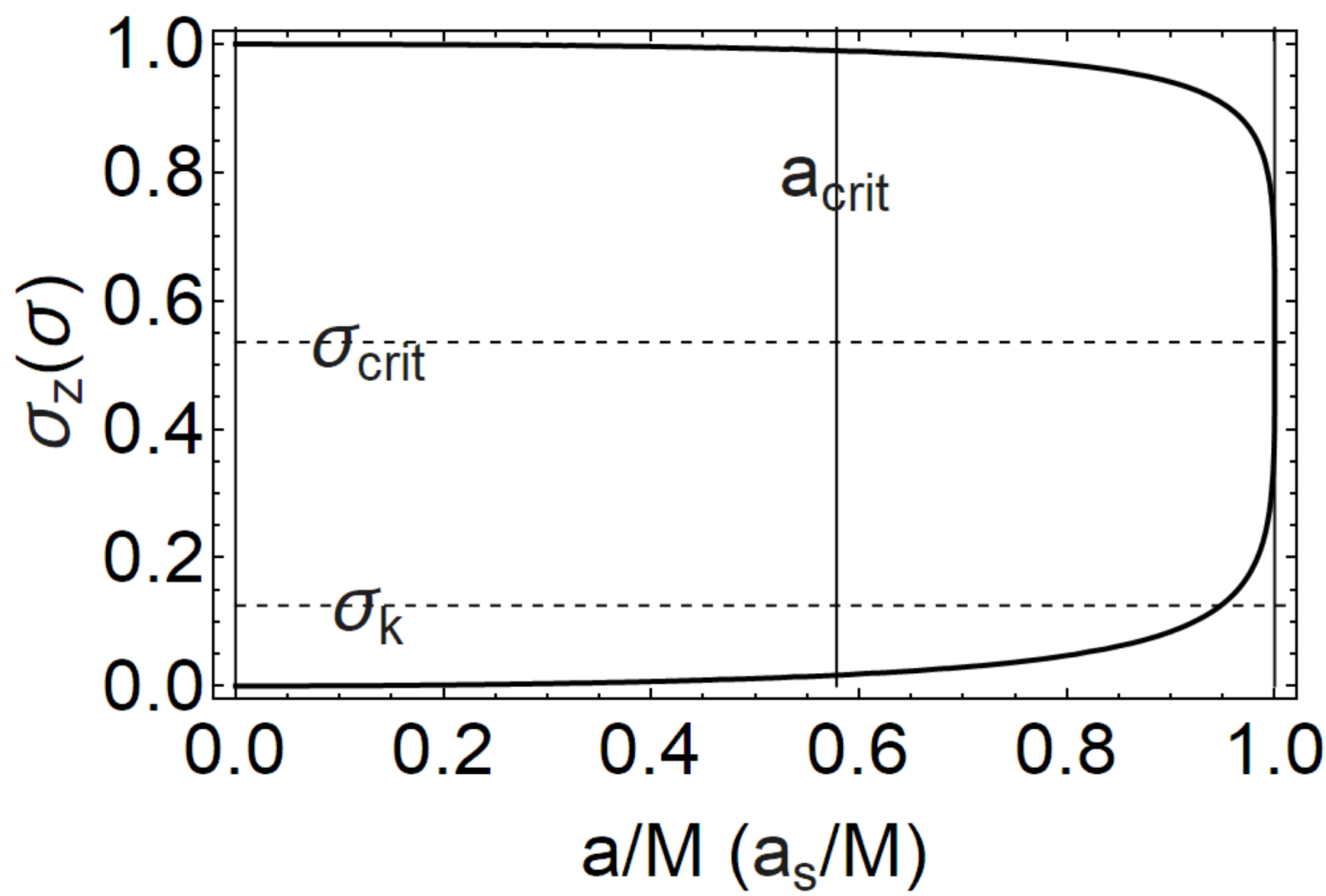}
  \caption{Upper line panels and  bottom left panel. Replicas analysis for  five different spin values. Plots show the horizons $r_{\pm}$ frequencies $\omega_H^{+}$ (red curve), $\omega_H^{-}$ (black curve), $-\omega_H^{+}$ (dashed red curve), and $-\omega_H^{-}$ (dashed black curve) on the plane $(\theta,r/M)$ for different values of the BH  horizon radius $r_H$. Here, the BH spin is $a_{\pm}=\sqrt{r_H(2-r_H)}$, $r_H\in ]0,M]$ is an inner horizon, and $r_H\in ]M,2M]$ is an outer horizon.
  The gray curves denote the ergosurfaces $r_{\epsilon}^{\pm}$. Inner and outer horizons correspond to vertical lines, with $r_H=M$ for the extreme case $a=M$.
  The inner horizon $r_H=0.16M$ corresponds to $a=0.542586M$, $r=r_{crit}$ corresponds to $a_{crit}\equiv 0.5784M$, $r_H=2/5 M$ to $a=0.8M$, and $r_H=M$ to $a=M$. The limiting planes $(\sigma_v,\sigma_z)$ are depicted as functions of the  dimensionless spin $a/M$. Alternatively, the plots show the limiting spins $(a_l, a_s)$ as functions of the plane $\sigma$--Eqs\il(\ref{Eq:sigmav},\ref{Eq:sigmaz},\ref{Eq:al},\ref{Eq:as}). The values $a_{crit}\equiv 0.5784M$,
 $\sigma_k=2 \left(8-3 \sqrt{7}\right)=0.125492$, and
  $\sigma_{crit}=2 \left(2-\sqrt{3}\right)\approx 0.536$,
  are shown explicitly.}\label{FIG:ARESE1Pola1}
\end{figure*}
%
%
\begin{figure*}\centering
   \includegraphics[width=5.6cm]{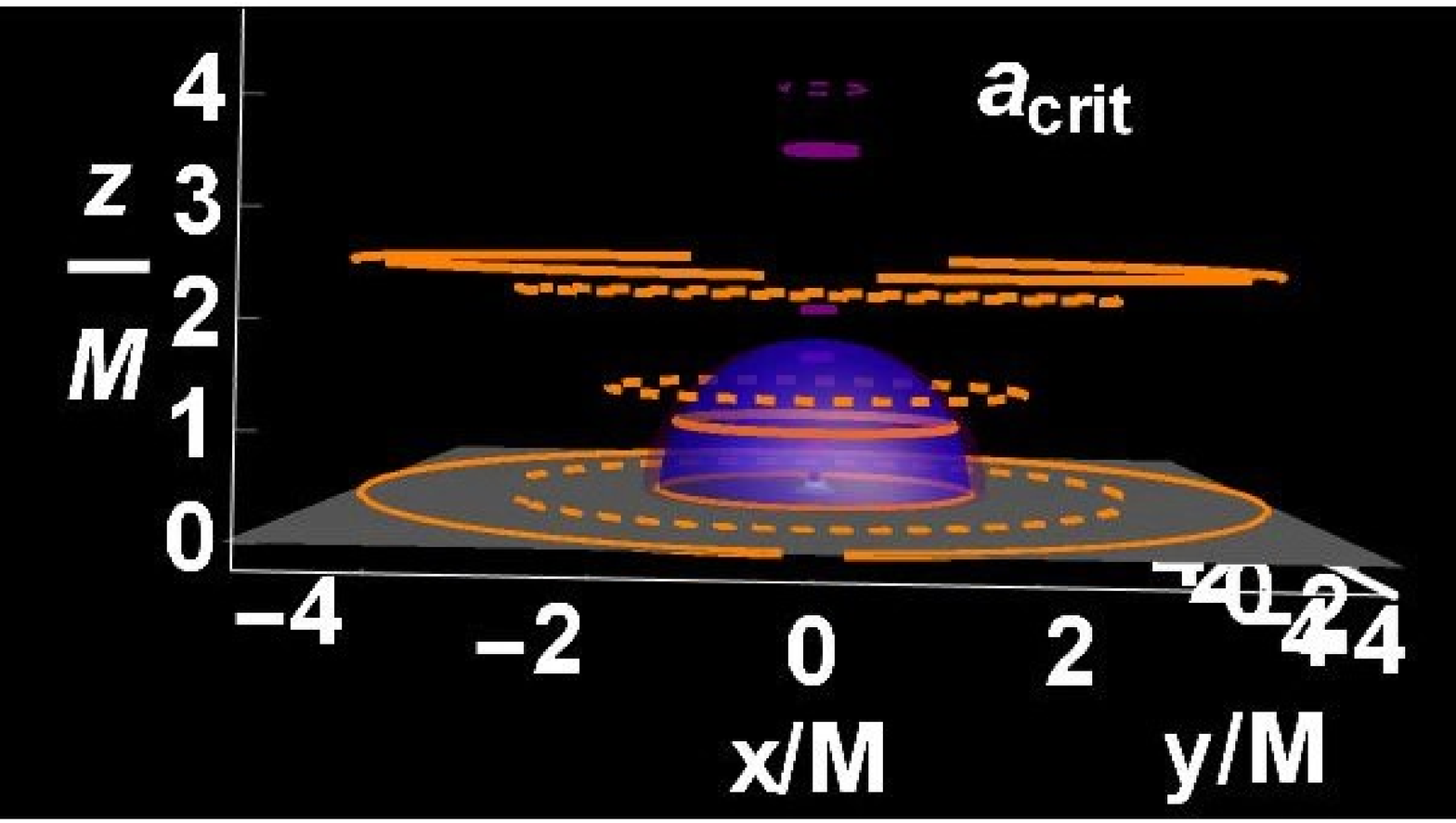}
  \includegraphics[width=5.6cm]{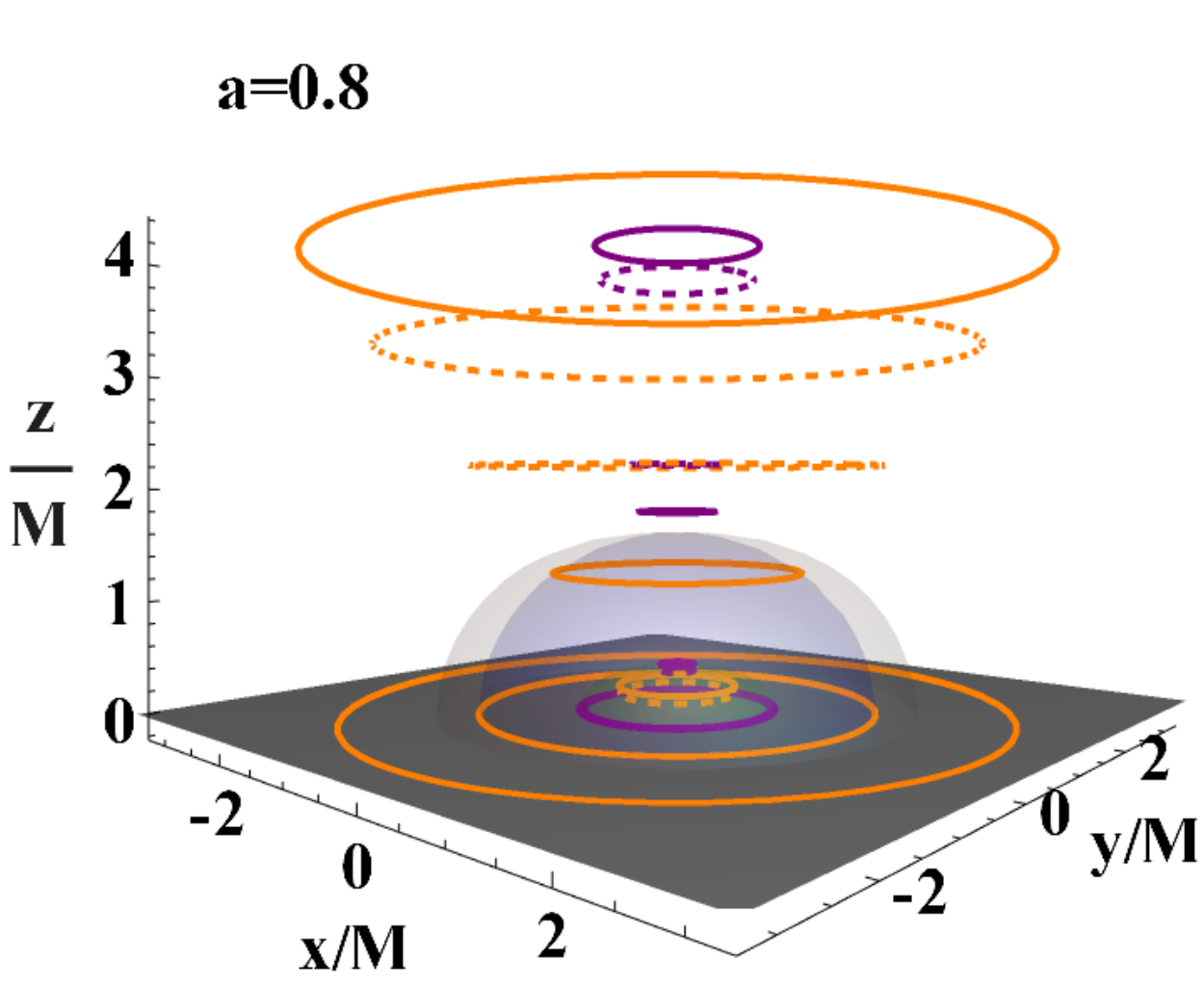}
  \includegraphics[width=6.5cm]{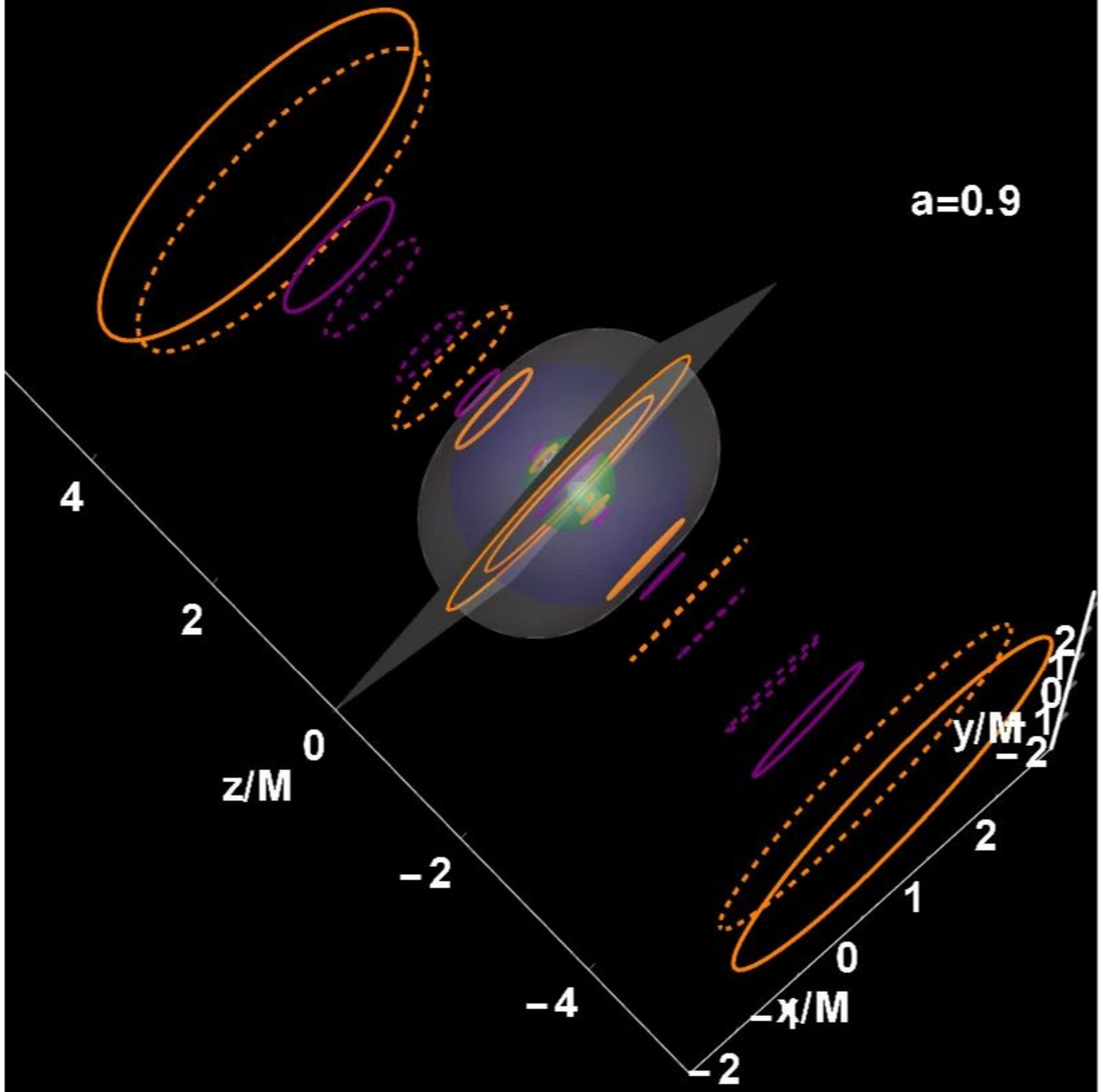}
   \caption{Examples of horizon replicas are  shown for  fixed  planes $\sigma\equiv \sin^2\theta$ and spins $a/M$. The
   green sphere represents the inner  horizon $r_-$, the blue sphere is the outer  horizon $r_+$, the light gray surfaces are the inner and outer ergosurfaces $r_{\epsilon}^{\mp}$. Orange orbits are replicas of the outer horizon frequency $\omega_H^{+}$, orange dashed curves are replicas of $-\omega_H^+$, purple curves  are replicas of $\omega_H^-$, and purple dashed curves are replicas of $-\omega_H^-$.
   In left and right  panels,  the coordinates are
    $\{x=r \sin\theta \cos\phi, y=r \sin\theta \sin\phi,z=r \cos\theta\}$. The singularity corresponds to the point $r=0$ at $x=0$, $y=0$, and $z=0$. In the central panel, the coordinates are $\left\{x=\sqrt{a^2+r^2} \sin \theta  \cos \phi,y=\sqrt{a^2+r^2} \sin\theta \sin\phi,z=r \cos \theta\right\}$. The singularity is the central disk of radius $a$. The spins are selected in agreement with the analysis of Fig.\il(\ref{FIG:ARESE1Pola1}). Here  $a_{crit}\equiv 0.5784M$. }\label{FIG:plot4sit7}
\end{figure*}
%
\section{Discussion }
\label{Sec:bundle-description}
The  replicas  analysis of Sec.\il(\ref{Sec:pri-photon-fre}) evidences how every photon circular frequency $\omega$ can be read as a BH horizon frequency, and in this sense  every photon circular orbit can be  read as  an horizon replica.
Replicas can therefore be grouped in structures, known as metric Killing bundles (MBs) containing  all (and only) the horizon replicas of every Kerr geometry,  for a given (photon) frequency  $\omega$, called   bundle characteristic frequency.  Therefore  MBs also   connect  measures in different  spacetimes, BH geometries  and  BH and naked singularities, grouping photon orbits in different geometries,   characterized by equal value of the photon orbital frequency  $\omega$. (Replicas in NSs were  interpreted   in \cite{remnants}
 as  ``horizons remnants'' and, more generally,   appear connected to the   "pre-horizon regime"  introduced in
\cite{de-Felice1-frirdtforstati,de-Felice-first-Kerr}.).
The  connection between replicas and MBs   appear evident  by the MBs definition: MBs with characteristic frequency $\omega$ are solutions of the condition
$a:\;g(\laa,\laa)=0$ for  the  fixed, constant {$\omega$}.
Therefore we can read the results of Sec.\il(\ref{Sec:pri-photon-fre}) in terms of   bundles properties.

MBs can be   represented as curves on a plane, known as  extended plane, tangent  to the curve $a_{\pm}\equiv \sqrt{r(2M-r)}$, representing all the BHs  horizons.
 The horizons curve $a_{\pm}>0$ is a functions of the horizons radii  $r/M$ in the extended plane, for  the inner BH  horizons, where  $r\in]0,M]$ (with $\omega=\omega_H^-\geq1/2$), or the  outer BH horizon, for $r\in[M,2M]$ (with $\omega=\omega_H^+\in ]0,1/2]$).
Introducing the quantify    $\la=a\sqrt{\sigma}$,
the extended plane of the  Kerr geometry
  is the plane $\la/M-r/M$,  or  $a/M-r/M$   in the special case of the equatorial plane. (Note that the  horizon curve  is obviously independent of the polar angle $\theta$.).
   Using the variable  $\la$  it is possible to  connect points in different planes $\sigma$,  as done in  Sec.\il(\ref{Sec:pri-photon-fre}).
 The significance of the MBs with respect to the Kerr horizons lies in the fact that,
   \textit{per}  construction,  each bundle  is a curve tangent to the horizon curve in the extended plane. Consequently  MB set  must necessarily contain at least one BH  geometry, and be tangent to  only one BH geometry  in the case of corotating  frequency   $\omega a>0$.
   For instance, the special  bundle with characteristic frequency $\omega=1/2$,  refers   to the extreme Kerr BH spacetime as its tangent point to  the horizon curve is  $(r=M, a=M)$  with  $\omega=\omega_H^\pm=1/2$  and, on the  equatorial plane, where the extended plane is $a-r$, the metric bundle set contains the  extreme Kerr BH   only  and NSs.
 In this sense,  remarkably,  the horizon curve in the extended  plane is generated as envelope surface of all the MBs curves.
The characteristic frequency of the bundles is  therefore in particular  the frequency of the BH horizon defined by the  tangency point, hence  each other point of the  MB curve  is its horizon replicas, and all the geometries of the curves are connected by this property.
Regarding the representation of  counter-rotating  horizons replicas described in Eqs\il(\ref{Eq:counter-mmm},\ref{Eq:counter-mmm1},\ref{Eq:counter-mmm2},\ref{Eq:counter-mmm3},
\ref{Eq:counter-mmm4},\ref{Eq:counter-mmm5},\ref{Eq:counter-mmm6},\ref{Eq:counter-mmm7},\ref{Eq:counter-mmm8},
\ref{Eq:counter-mmm9},\ref{Eq:counter-mmm10}) in the extended plane, we note that, by  using the symmetries of the   $\mathcal{L}$  tensor and the  Kerr metric  tensor  discussed in  Sec.\il(\ref{Sec:pri-photon-fre}), we can represent counter-rotating replicas in a compact and immediate way by  extending the extended plane into the negative region, with   $\mathcal{A}<0$ (equivalently  $a<0$). In this representation, the  bundle characteristic frequency is always positive, and the  curves extend,   mostly continuously,  in the  negative section of the extended plane, at $a<0$, grouping   corotating replicas of the horizon frequency  in the positive sector $a>0$, and horizon counter-rotating replicas in the extension $a<0$. In this extension,  the horizon curve is a circle of  radius  $M$ centered on the point $(a=0, r=M)$.
Per construction the region  ($a^2<a^2_{\pm}$ or $r\in ]r_-,r_+[$),  bounded   by the horizon curve,  is inaccessible  for  the bundles which  are not defined in this region.
In this context, the inaccessibility of some frequencies for the observer  (as for a  part  of the  inner  horizon curve frequencies), as  concept of  horizon confinement highlighted in Sec.\il(\ref{Sec:pri-photon-fre}), can be expressed in the  MBs  frame  as   ``local  causal ball'', that is a  entirely confined region,    inaccessible  to  any other   region of the extended plane.
Although we also consider also   the  inner region ($r<r_-$), in this analysis we are  mostly interested in the  information contained in the outer region  $r>r_+$ and therefore accessible to the observer. The bundle curves with no portion in the outer region, are confined and the information is unaccessible for  the observer.

Significantly  then we can interpret   results of Sec.\il(\ref{Sec:pri-photon-fre})  in the     metric bundles framework.
Replicas connect the   inner region of the extended plane, bounded by  the  inner  horizon,  to  the region  outside  the outer horizon.
In this context we can explain  how an  observer can  eventually   measure  a replica $\omega_{\circledast}$ of a BH horizon  with  spin  $a_{\circledast}$ on the orbit  $q_{\circledast}$.  The point  $q_{\circledast}$ belongs  to the   MB with characteristic frequency $\omega_{\circledast}$ which, in the observer spacetime $(a=a_{\circledast})$, crosses the point $q_{\circledast}$ and it is tangent to the observer BH horizon curve in the extended plane to the point  ($a_{\circledast}$, $r_{\circledast}$).  The replica registered by the  observer is, as discussed in Sec.\il(\ref{Sec:pri-photon-fre}),  the   BH corotating or counter-rotating outer  horizon frequency $\omega_{\circledast}=\pm\omega_H^+$  or  the  inner horizon frequency  $\omega_{\circledast}=\pm\omega_H^-$.  The stationary observer located at $q_{\circledast}$  has  orbital frequency     $\omega_o\in]\omega_{1},\omega_{2}[$, where $(\omega_{1},\omega_{2})$ are the limiting photon frequencies $\omega_{\pm}$, and  one of the  frequencies $(\omega_{1},\omega_{2})$  is the  frequency $\omega_{\circledast}$ of the BH inner $r_{\circledast}=r_-(a_\circledast)$ or outer horizon $r_{\circledast}=r_+(a_{\circledast})$,  which is replicated on $q_{\circledast}$.   The second  light-like frequency of the pair $(\omega_{1},\omega_{2})=\omega_{\pm}$   is, according to Sec.\il(\ref{Sec:pri-photon-fre}), an horizon  replica of a different  BH  with spin $a_{\odot}\neq a_{\circledast}$, included in  the metric bundle with that  characteristic frequency and tangent to the inner or outer BH horizon with spin $a_{\odot}$.
The relation between the two  frequencies $(\omega_{1},\omega_{2})=\omega_{\pm}$ on  $q_{\circledast}$ is determined by a characteristic ratio, relating therefore also the BHs with spins $(a_{\odot}, a_{\circledast})$.
The two  metric bundles with characteristic frequencies  $(\omega_{1},\omega_{2})$ respectively cross on $q_{\circledast}$.
 (For a point $q_{\circledast}$ there  is a  maximum of two crossing bundles.).

 We can  express more precisely and in detail  the relation between the replicas and bundle structure, with respect to the analysis  of  Sec.\il(\ref{Sec:pri-photon-fre}). The  bundle curves have two notable points. In the extended plane the bundle  has  a tangent point  $(a_g, r_g)$ on the horizon curve, defining the characteristic frequency of the bundle as horizon frequency. However, the bundle has also an origin, corresponding to the point at $r=0$ (the singularity limit in this  frame). The horizon frequency is the characteristic frequency of the bundle and  therefore the frequency at each point of the bundle,  there is then at the bundle  origin  $a_0$ $(\la_0)$ for  $r=0$, defined as  $a_0\equiv1/\omega\sqrt{\sigma}$ {(or $\la_0=a_0\sqrt{\sigma}=1/\omega$)} .
Replicas of a frequency $\omega$, studied  in  Sec.\il(\ref{Sec:pri-photon-fre})  in a fixed BH spacetime with spin  $\bar{a}$, is provided by the point of the  bundle crossing the horizontal lines $a=\bar{a}=$constant (or $\la=$constant, considering explicitly the dependence on the plane $\sigma$). Therefore it is a quantity dependent on the  bundles curvature in the extended plane, as solutions  of  $a(\omega)=\bar{a}$, being $a(\omega)$ the bundles with characteristic frequency $\omega$.

Below we  consider  explicitly the relation between the frequency replicas, the bundles tangency condition to the horizon curve in the extended plane and the bundles origin spin.
According to the value of  $\sigma$, the inner horizon can be generated  (tangent to the MB) either by MBs with  naked singularities origins or   with  BH origins. A MB with
BH origin generates only the inner  horizon,  this is because there is  $\omega\equiv 1/\la_0\equiv 1/(a_0\sqrt{\sigma})\geq1/a_0\geq1$ (in magnitude) and therefore the relative bundle characteristic frequency is an inner horizon frequency (it is indeed a section of  inner horizon curve in the extended plane, as the inner horizon frequencies are $\omega>1/2$, the case $\omega=1$ is therefore a discriminant case  with $a_g/M =
    4/5, r_ +/M= 8/5,
r_ -/M= 2/5$.). Or specifically we can summarize the relations  between $(\omega, a_0, \sigma)$ as follows:
\bea
&&\mbox{for}\quad \omega\in ]0,1/2[\quad \mbox{there is }\quad  a_0>2,
\\
&&\mbox{for}\quad\omega=1/2\quad \mbox{there is }\quad  a_0\geq 2,
\\&&\nonumber
\mbox{for}\quad\omega\in ]1/2,1[\quad \mbox{there is}
\\
&&\nonumber\quad (a_0\in ]1,2[, \sigma\in ]1/a_0^2,1])
;(a_0\geq2, \sigma\in ]1/a_0^2,4/a_0^2[),
\\
&&\nonumber
\mbox{for}\quad\omega>1  \quad \mbox{there is }
\\
&&\nonumber\quad (a_0\in [0,1],\sigma\in ]1/\omega^2,1]);(a_0>1,
 \sigma \in ]0,1/\omega^2[).
\eea
%
 %
  On the other hand,  as seen in Sec.\il(\ref{Sec:pri-photon-fre}) there are orbits with characteristic  frequencies equal to  the outer horizon  frequencies,  which are located in the inner
region of the extended plane.
A limiting special  case is represented by the equatorial plane, $\sigma=1$, where the extended plane is $a-r$ and the bundles origin is $a_0=1/\omega$. This relevant case has been analyzed  in Eqs\il((\ref{Eq:invt},\ref{Eq:counter-mmm6},\ref{Eq:counter-mmm7},\ref{Eq:counter-mmm7a}). We evidenced how  MBs  with BH  origins are  tangent to the inner horizon (for any plane $\sigma$), however for  frequencies $\omega=\omega_H^-\geq1$ with
 $a_g\in[0,0.8M]$ and tangent radius $r_g\in[0,2/5M]$,  MBs are confined in  the region of the extended plane upper-bounded  by the inner horizon.
Thus, more generally, for large values of $\sigma\in[0,1]$, and particularly on the equatorial plane MBs with BH origin spin $a_0$ are  {confined} in the inner region of the extended plane. Consequently, on these planes $\sigma$s,
all the  frequencies   $\omega\geq1$  defining  these bundles  are confined and cannot be found in the outer region, i.e. $r>r_+$.
Therefore, the  confinement of  the  MBs tangent to  the inner horizons  can be overcomed, as noted in Sec.\il(\ref{Sec:pri-photon-fre}),
by considering that
the
zero-quantity  $g(\mathcal{L},\mathcal{L})$   depends explicitly on the plane $\sigma>0$ (and not from the bundle origin $\la_0$ only).
For each point  of the horizons curve, it is possible to find a  replica for a plane  $\sigma$  bounded  in $\sigma<\sigma_{crit}$. 
This  implies that  close to the rotational axis, $\sigma\ll1$,  for a spacetime with spin $a$ it is possible to find an inner horizon replica  in the outer region  $r>r_+$.
 Consequently,  bundles can extract ``information'' on the inner horizons frequencies  near the rotation axis and, therefore, in this sense, the  inner region is not entirely confined.  Another significant  aspect of the horizon confinement  concerns the intriguing possibility of extracting  information from counter-rotating orbits, while
 from the observational view-point, we established that the rotational axis of a Kerr BH  may contain important information about the singularity and horizons.

\section*{Acknowledgements}
	This work was partially supported  by UNAM-DGAPA-PAPIIT, Grant No. 114520,
	Conacyt-Mexico, Grant No. A1-S-31269,
	and by the Ministry of Education and Science  of Kazakhstan, Grant No.
	BR05236730 and AP05133630.

\appendix
\section{Solutions}
{\footnotesize
\bea&&\label{Eq:rn}
r_n:\sum_{i\equiv 0}^8 n_i r^i\equiv 0,\quad\mbox{where}
\\\nonumber
&&n_0\equiv a^6 (\sigma -1)^2 \left[(a^4 \sigma ^2+8 \left(a^2-2\right) \sigma +16\right];
\\\nonumber
&& n_1\equiv -4 a^4 (\sigma -1) \left[a^2 \sigma  \left(\sigma  \left[a^2 \sigma +8\right]-4\right)-8 \sigma  (\sigma +1)-16\right];
\\\nonumber
&&n_2\equiv 2 \left[a^8 (\sigma -2) (\sigma -1) \sigma ^2+2 a^6 \sigma  [(\sigma -2) \sigma  (\sigma +4)+6]+\right.
		\\
		&&\nonumber\qquad \left. 8 a^4 [\sigma  (\sigma  (\sigma +6)-5)+2]+32 a^2 (\sigma +1)^2\right];
\\\nonumber
&&n_3\equiv -4 \left[a^6 (\sigma -2) \sigma ^3+16 a^2 \left(\sigma ^2+1\right)+4 a^4 \sigma  ((\sigma -4) \sigma +2)\right];
\\\nonumber
&&n_4\equiv a^2 \left[\sigma  \left(a^4 \sigma  [(\sigma -6) \sigma +6]+8 a^2 (3-2 \sigma )+32 \sigma -48\right)+16\right];
\\\nonumber
&&n_5\equiv 4 \sigma  \left[a^4 \sigma ^2+4 a^2 (\sigma -1)+8\right];
\\\nonumber
&&n_6\equiv -2 \sigma  \left[a^4 (\sigma -2) \sigma -4 a^2+8\right];\quad  n_7\equiv 0
\\\nonumber
&&n_8\equiv a^2 \sigma ^2
\eea
\bea
&&\label{Eq:rz}
r_z: \sum_{i\equiv 0}^6 z_i r^i\equiv 0
\\\nonumber
&&
z_0\equiv a^4 (\sigma -1)^2 \left[a^4 \sigma ^2+8 \left(a^2-2\right) \sigma +16\right];
\\\nonumber && z_1\equiv -2 a^2 (\sigma -1) \left[a^2 \sigma -4\right] \left[\sigma  \left(a^2 (\sigma +1)-4\right)+4\right];\\
&&\nonumber
z_2\equiv a^2 \left[\sigma  \left[a^4 (\sigma -3) (\sigma -1) \sigma +4 a^2 (4-3 \sigma )+16 (\sigma -2)\right]+16\right];\\
&&\nonumber z_3\equiv 4 a^2 \left(a^2-2\right) \sigma ^2;\\
&&\nonumber
z_4\equiv \sigma  \left[a^4 \sigma  (3-2 \sigma)+4 a^2 (\sigma +2)-16\right];\quad z_5\equiv 2 a^2 \sigma ^2;\quad z_6\equiv a^2 \sigma ^2.
\eea}
{\footnotesize
\bea&&\label{Eq:as}
a_s\equiv \sum_{i\equiv 0}^{10}c_i a^i\equiv 0,\quad  i\equiv even
\\\nonumber
&&
c_0\equiv -6912 (\sigma -1)^2 \sigma;\\&&\nonumber
c_2\equiv 16 (\sigma -1) [\sigma  (\sigma  (97 \sigma +447)-312)-16]; \\\nonumber
&&c_4\equiv 8 \sigma  [\sigma  (\sigma  ([1045-411 \sigma] \sigma -1228)+588)+16];
\\&&\nonumber  c_6\equiv \sigma ^2 [\sigma  (\sigma  [\sigma  (201 \sigma +1160)-2784]+1408)+16];
\\&&\nonumber
c_8\equiv -2 (\sigma -1) \sigma ^4 [3 \sigma  (5 \sigma -28)+68];
\quad
c_{10}\equiv (\sigma -1)^2 \sigma ^6;
\eea}
{\footnotesize
\bea
&&\label{Eq:sigmav}
\sigma_v:\sum_{i\equiv 0}^{12}v_i \sigma^i\equiv 0
\\&&\nonumber
v_0\equiv 4096 a^2 \left(a^2-1\right)^2;
\\&&\nonumber
v_1\equiv 2048 (a-1) (a+1) \left(a^6+16 a^4-87 a^2+54\right);
\\&&\nonumber
v_{2}\equiv 256 \left(a^{10}+220 a^8-466 a^6+972 a^4+1257 a^2-1728\right);
\\&&\nonumber
v_3\equiv 512 \left(41 a^{10}+431 a^8-823 a^6-5155 a^4+4754 a^2-1296\right);
\\&&\nonumber
v_4\equiv 256 \left(9 a^{12}+577 a^{10}-5232 a^8+18447 a^6+961 a^4+3606 a^2-1728\right)
\\&&\nonumber
v_5\equiv 256 \left(98 a^{12}-2083 a^{10}+10317 a^8-20461 a^6+\right.
\\
&&\nonumber\qquad \left. 6741 a^4+2748 a^2-432\right);
\\&&\nonumber
v_6\equiv 32 a^2 \left(35 a^{12}-2804 a^{10}+19730 a^8-69292 a^6+
\right. \\
&&\nonumber \qquad \left. 64139 a^4+51992 a^2+2504\right);
\\\nonumber
&&v_7
\equiv -64 a^4 \left(72 a^{10}-1541 a^8+6823 a^6-8371 a^4+14253 a^2-8164\right);
\\&&\nonumber
v_8\equiv 16 a^6 \left(9 a^{10}+417 a^8-2288 a^6+13455 a^4+4481 a^2+566\right);
\\&&\nonumber
v_9\equiv -8 a^8 \left(45 a^8+497 a^6-29 a^4+5507 a^2-3972\right);
\\&&\nonumber
v_{10}\equiv a^{10} \left(a^8+284 a^6+686 a^4+204 a^2-919\right);
\\&&\nonumber
v_{11}\equiv -2 a^{12} \left(a^6+33 a^4-53 a^2+19\right)
\\&&\nonumber
v_{12}\equiv a^{14} \left(a^2-1\right)^2;
\eea}{\small \bea&&\label{Eq:sigmaz}
\sigma_z: \sum_{i\equiv 0}^8 c_i \sigma^i\equiv 0
\\\nonumber
&&
x_0\equiv 256 a^2;\quad x_1\equiv 128 \left(a^4+37 a^2-54\right)
,\\&&\nonumber  x_2\equiv 16 \left(a^6+294 a^4-759 a^2+864\right);
\\&&\nonumber
x_3\equiv 32 \left(44 a^6-307 a^4+175 a^2-216\right);
\\&&\nonumber
 x_4\equiv 8 a^2 \left(17 a^6-348 a^4+1045 a^2+194\right);
\\&&\nonumber
x_5\equiv -8 a^4 \left(38 a^4-145 a^2+411\right);
\\&&\nonumber
x_6\equiv a^6 \left(a^4+198 a^2+201\right);
\quad
x_7\equiv -2 a^8 \left(a^2+15\right);
\quad x_8\equiv a^{10};
\eea}
{\footnotesize \bea
&&\label{Eq:al}
a_l: \sum_{i\equiv 0}^{18}l_i a^i,\quad i\equiv even
\\\nonumber
&&
l_0\equiv -110592 \sigma  (\sigma +1)^4;
\\&&\nonumber
l_2\equiv 256 [\sigma  (\sigma  [\sigma  (\sigma  [\sigma  (313 \sigma +2748)+3606]\\
&&\nonumber \qquad +9508)+1257]+1128)+16];
\\\nonumber
&&l_4\equiv 256 [\sigma  (\sigma  [\sigma  (\sigma  [\sigma  (\sigma  [2041 \sigma +6499]+6741)+
\\
&&\nonumber \qquad \qquad961]-10310)+972]-824)-32];
\\\nonumber
&&l_6\equiv 32 [\sigma  (\sigma  [\sigma  (\sigma  [\sigma  (\sigma  [\sigma  (283 \sigma -28506)+64139]+\\
&&\nonumber\qquad -163688)+147576]-13168)-3728]
+960)+128];
\\\nonumber
&&l_8\equiv 16 \sigma  [\sigma  (\sigma  [\sigma  (\sigma  [\sigma  (\sigma  [\sigma  (1986 \sigma +4481)+33484]-
\\
&&\nonumber \quad 138584)+165072]-83712)+13792]
+3520)+128];
\\\nonumber
&& l_{10}\equiv \sigma ^2 [\sigma  (\sigma  [147712-\sigma  (\sigma  [\sigma  (\sigma  [\sigma  (919 \sigma +44056)\\
&&\nonumber\qquad -215280]+436672)-631360]+533248)]+20992)+256]
\\&&\nonumber
l_{12}\equiv 2 \sigma ^4 [\sigma  (\sigma  [\sigma  (\sigma  [\sigma  (\sigma  [102-19 \sigma]+116)\\
&&\nonumber\qquad -18304]+49312)-44864]+12544)+1152
]\\\nonumber&&l_{14}\equiv \sigma ^6 [\sigma  (\sigma  [\sigma  (\sigma  [\sigma  (\sigma +106)+686]
\\
&&\nonumber\qquad \qquad -3976)+6672]-4608)+1120];
\\\nonumber
&&l_{16}\equiv -2 (\sigma -2) (\sigma -1) \sigma ^8 [\sigma  (\sigma +36)-36];
\\&&\nonumber
l_{18}\equiv (\sigma -1)^2 \sigma ^{10}.
\eea}
%
%

\end{document}